\documentclass[aip, preprint]{revtex4-1}

\usepackage{amsmath}
\usepackage{graphicx}
\usepackage{textcomp}

\usepackage{hyperref, color}
\definecolor{linkColor}{rgb}{0.8,0,0}
\definecolor{darkred}{rgb}{0.8,0,0}
\hypersetup{pdfborder={0 0 0},colorlinks=true,urlcolor=linkColor,citecolor=linkColor}

\usepackage{fancyhdr}
\begin{document}

\title{An Automated Scanning Transmission Electron Microscope Guided by Sparse Data Analytics}

\author{Matthew Olszta}
\thanks{These authors contributed equally.}
\affiliation{Energy and Environment Directorate, Pacific Northwest National Laboratory, Richland, Washington 99352}

\author{Derek Hopkins}
\thanks{These authors contributed equally.}
\affiliation{Environmental Molecular Sciences Laboratory, Pacific Northwest National Laboratory, Richland, Washington 99352}

\author{Kevin R. Fiedler}
\thanks{These authors contributed equally.}
\affiliation{College of Arts and Sciences, Washington State University -- Tri-Cities, Richland, Washington 99354}

\author{Marjolein Oostrom}
\affiliation{National Security Directorate, Pacific Northwest National Laboratory, Richland, Washington 99352}

\author{Sarah Akers}
\affiliation{National Security Directorate, Pacific Northwest National Laboratory, Richland, Washington 99352}

\author{Steven R. Spurgeon}
\email{steven.spurgeon@pnnl.gov}
\affiliation{Energy and Environment Directorate, Pacific Northwest National Laboratory, Richland, Washington 99352}

\date{\today}


\begin{abstract}

Artificial intelligence (AI) promises to reshape scientific inquiry and enable breakthrough discoveries in areas such as energy storage, quantum computing, and biomedicine. Scanning transmission electron microscopy (STEM), a cornerstone of the study of chemical and materials systems, stands to benefit greatly from AI-driven automation. However, present barriers to low-level instrument control, as well as generalizable and interpretable feature detection, make truly automated microscopy impractical. Here, we discuss the design of a closed-loop instrument control platform guided by emerging sparse data analytics. We demonstrate how a centralized controller, informed by machine learning combining limited \textit{a priori} knowledge and task-based discrimination, can drive on-the-fly experimental decision-making. This platform unlocks practical, automated analysis of a variety of material features, enabling new high-throughput and statistical studies.

\end{abstract}

\maketitle

\section{Introduction}

The history of science is punctuated by the development of tools, approaches, and protocols to more richly probe the natural world.\cite{Daston2011} Watershed discoveries have been directly linked to humanity's sophistication in designing and executing increasingly revealing experiments. The rise of clean energy,\cite{Zhang2019} the silicon revolution,\cite{VanBenthem2007} and designer medicine\cite{Shen2018,Frank2017} are just a few results of seeing the world through better spatial, chemical, and temporal lenses. Traditionally, manual approaches have kept pace with experimentation, but today all scientific domains produce data at a scale and complexity far exceeding human cognition.\cite{Gupta2018, Baraniuk2011} This situation has yielded an ironic surplus of data and shortfall of immediately actionable knowledge, motivating the development of automation and artificial intelligence (AI) to transform experimentation.\cite{Stach2021,Noack2021,Brown2020,Vasudevan2019, King2004} While some communities, such as chemical synthesis,\cite{Xu2020,Shields2021,Higgins2020,Hase2019,Takeuchi2005} crystallography,\cite{Noack2021,Arzt2005,Abola2000} and biology,\cite{Shen2018,Nogales2015,King2004,Carragher2000} were early adopters of this paradigm, fields such as electron microscopy of hard matter have just begun this transition because of longstanding practical barriers.\cite{Hattar2021,Ede2020,Schorb2019,Kalinin2015} Some of these barriers, such as closed or proprietary instrumentation platforms, are the result of business drivers, while others stem from a lack of accessible, standards-based experiment frameworks.\cite{Spurgeon2020c, Kalidindi2015} As a result, the adoption of data science in microscopy has been highly fragmented, with some institutions able to develop powerful custom instrumentation and analysis platforms, while others have been unable to integrate these practices into everyday analysis workflows.\cite{MaiaChagas2018} There is presently a great need to design a practical and generalizable automation platform to address common use cases.

The two essential components of any automation platform are \textit{low-level instrument control} and \textit{decision-making analytics}. In the case of the former, researchers are typically forced to choose between accepting the control limitations set by manufacturers (often necessary to guarantee performance specifications) or designing a bespoke instrument. The community has developed multiple innovative approaches to high-throughput screening and automation,\cite{Schorb2019,Mastronarde2003a,Carragher2000} most prominently in the fields of cell biology\cite{Yin2020, Coudray2011}, medical diagnostics,\cite{Martin-Isla2020,Gurcan2009} single-particle cryo electron microscopy,\cite{Liu2016a, Frank2017} but also in crystallography,\cite{Rauch2014} semiconductor metrology,\cite{Strauss2013} and particle analysis.\cite{Uusimaeki2019,House2017} More recently, manufacturers have begun to provide increased access to low-level instrument functions.\cite{Kalinin2021,PyJEM,Meyer2019a} These application programming interfaces (APIs) can potentially be integrated into existing machine learning (ML) pipelines\cite{Ziatdinov2021} and may enable new control frameworks, such as ``measure-by-wire'' auto-tuning\cite{Tejada2011} and Gaussian-process-driven experimentation.\cite{Stach2021, Vasudevan2019} However, because these APIs are in their incipient development phase and require programming, hardware, and microscopy expertise, few control systems have been designed to take advantage of them. Further complicating the situation, modern instruments often incorporate components from different manufacturers (e.g., cameras, spectrometers, and scan generators), whose lack of feature and access parity complicate any ``open controller'' design. The ideal ``open controller'' should (1) serve as a central communications hub for low-level instrument commands, (2) scale to include additional hardware components, (3) connect to external sources of data archival, and (4) integrate on-the-fly feedback from analytics into the control loop. Ultimately, the goal of any such system is to handle low-level commands, allowing the researcher to focus on high-level experiment design and execution. This goal does not mean that the researcher should be ignorant of the underlying operation of the instrument; rather, it acknowledges that most experiments aim to collect physically meaningful materials and chemical descriptors (e.g., morphology, texture, and local density of states), \cite{Curtarolo2013} rather than raw (meta)data information (e.g., stage coordinates, probe current, and detector counts).

Alongside practical low-level instrument control, decision-making analytics is another necessary part of any automation platform. Traditional instrument operation has been based on human-in-the-loop control, in which a skilled operator manually defines experimental parameters, collects data, and evaluates outputs to decide next steps. However, this approach is poorly suited to the large data volumes and types now routinely generated;\cite{Spurgeon2020c} humans have trouble analyzing higher dimensional parameter spaces, are prone to bias and omission of steps, and often cannot respond fast enough.\cite{Taheri2016} Analytics approaches must therefore be developed that can quickly define actionable metrics for closed-loop control. The field of computational imaging has devised approaches\cite{Voyles2016} to both improve data quality (e.g., denoising and distortion correction), and extract information using methods such as component analysis and ML. Deep learning approaches, such as convolutional neural networks (CNNs), have grown in popularity in microscopy because of their ability to learn generalizable models for trends in data without specific \textit{a priori} knowledge of underlying physics.\cite{Roccapriore2021,Ede2020} These methods can effectively interrogate large volumes of data across modalities\cite{Belianinov2015} and can be accelerated using embedding computing hardware to reduce processing times. Among its many applications, ML has been used to effectively quantify and track atomic-scale structural motifs\cite{Madsen2018a,Ziatdinov2017}, and has shown recent successes as part of automated microscope platforms.\cite{Trentino2021, Kalinin2021} Despite these benefits, CNNs are inherently constrained, since they typically require large volumes ($100$ to $>10$k images) of tediously hand-labeled or simulated training data.\cite{Aguiar2020} Due to the wide variety of experiments and systems studied in the microscope, such data is often time-consuming or impossible to acquire. In addition, a training set is typically selected with a predetermined task in mind, which is difficult to change on-the-fly to incorporate new insights obtained during an experiment.

Recently, few-shot ML has been proposed as one alternative approach to learn novel visual concepts based on little to no prior information about data.\cite{Rutter2019, Finn2017, Altae-Tran2017} Few-shot is part of the broader field of sparse data analytics, which targets the challenge of learning with limited data.\cite{Yao2021} In this approach, offline CNN pre-training is performed once using a typical network, such as Resnet101,\cite{He2016b} followed by the online application of a meta-learner specialized using a limited number of user-provided examples; this approach has the benefit of computational efficiency (since the offline training is performed only once) and flexibility to adapt to different tasks. Few-shot has seen minimal usage within the materials science community, primarily in the analysis of electron backscatter diffraction (EBSD) patterns,\cite{Kaufmann2021} but it has great potential to inform triaging and classification tasks in novel scenarios. We have recently demonstrated the efficacy and flexibility of the few-shot approach for segmentation of electron microscope images;\cite{Akers2021} using just 2--10 user-provided examples in an intuitive graphical user interface (GUI),\cite{Doty2021} it is possible to quickly classify microstructural features in both atomic-resolution and lower-magnification images. The output of the few-shot approach is essentially a feature map and statistics on their relative abundance. In addition, it is possible to easily extract pixel coordinates for desired features, which offers a pathway to feedback in a closed-loop automation system.

Here we describe the design of a scanning transmission electron microscope (STEM) automation platform based on closed-loop feedback provided by few-shot ML. We demonstrate the ability to acquire data automatically according to a predefined search pattern through a central instrument controller. This data is passed to an asynchronous communication relay, where it is processed by a separate few-shot application based on user input. The processed data is used to identify desired features and guide the subsequent steps of the experiment. Additionally, we demonstrate that, in combination with a stage montaging algorithm, automated data collection can be performed over large regions of interest (ROIs). A particular advantage of the few-shot approach is that it can classify features and guide the system by \textit{task}, which can be changed on-the-fly as new knowledge is gained. We demonstrate how this approach can lead to more intelligent and statistical experimentation in both open- and closed-loop acquisition scenarios.

\section{Materials and Methods}

\subsection{Hardware}

The microscope used in this study is a probe-corrected JEOL GrandARM-300F STEM equipped with the PyJEM Python API. The data shown is acquired in STEM mode at 300 kV accelerating voltage at magnifications ranging from 20--25 kx. Data processing is performed on a separate remote Dell Precision T5820 Workstation equipped with a Intel Xeon W-2102 2.9GHz processor 
and 1GB NVIDIA Quadro NVS 310 GPU.

\subsection{Automation Software}

The automation system is composed of interlinked hardware-software components. HubEM acts as the main end-use application for the system. It serves as a point for entering configuration, storing data, and directing the cooperation of other components through inter-process communication. It is implemented in C\#/Python and uses Python.NET 2.5.0, a library that allows Python scripts to be called from within a .NET application. PyJEM Wrapper is an application that wraps the PyJEM 1.0.2 Python library, allowing communication to the TEMCenter control application from JEOL. It is written in Python and runs on the JEOL PC used to control the instrument. GMS Python allows communication to the Gatan Microscopy Suite (GMS) 3.4.3. It runs as a Python script in the GMS embedded scripting engine. All components communicate using a protocol based on ZeroMQ and implemented in PyZMQ 19.0.2.

\subsection{Few-Shot Machine Learning}

The application for few-shot ML analysis has been described elsewhere.\cite{Doty2021} In brief, the application integrates Python, D3, JavaScript, HTML/CSS, and Vega-lite with Flask, a Python web framework. The front-end interactive visualization was created with JavaScript and HTML/CSS. The Flask Framework allows the inputs from the front-end user interaction to be passed as input to the Python scripts on the back-end. The Python scripts include the few-shot code\cite{Akers2021} for processing the image and the WizEM code for receiving the image and sending back the processed image.

\subsection{Image Stitching}

Stitching is performed using a custom Python 3.7.1 script. It can be run locally as a library or run as a stand-alone application on a remote machine to gain more processing power. The script works as follows. First, the acquired images were converted to grayscale to remove redundant information, as all the RGB channels are identical. Then the images were normalized to have mean pixel intensity of 0 and the maximum of the absolute value of intensity was normalized to 1 to adjust for differences in illumination or contrast. The cross correlation of the two images was then computed for every possible overlap between them. While it is not computed this way for efficiency purposes, intuitively the cross correlation can be thought of as sliding one image over the other and pointwise multiplying the pixel values of the overlapping points and then summed. Larger values of the cross correlation correspond to better agreement in the features of the two images because similar values, either positive or negative, will square to positive contributions. If the values are dissimilar (i.e., a mixture of positive and negative values) then they tend to cancel out which leads to smaller values that indicate worse agreement. As such, this computed value is used to search the possible overlaps to find a local maximum. However, there is subtlety in how this maximum is determined, because if the image shows periodic structure, then there could be many local minima in the cross correlation. In general, the global maximum will typically occur when the images are almost completely overlaid because there are numerous pixels that are summed, even if the overall alignment is poor. To compensate for this effect, and to emphasize the fact that we are prioritizing alignment of images, the cross correlation is normalized by the number of pixels that were summed to compute the value.

\clearpage

\section{Results and Discussion}

The design of any external microscope control system is naturally complex, since hardware components from multiple vendors must be networked to a custom controller and analysis applications. For simplicity, we divide our design into three systems: \textit{Operation}, \textit{Control}, and \textit{Data Processing}. The Operation System is a communication network that connects complex, low-level hardware commands to a simple, high-level user interface. The Control System encompasses both open- and closed-loop data acquisition modes, based on few-shot ML feature classification. Where the Operation System abstracts hardware commands, the Control System abstracts raw data into physically meaningful control set points. We demonstrate the operation of such a control scheme in the context of a statistical analysis of MoO$_3$ nanoparticles. Finally, the Data Processing System includes on-the-fly and \textit{post hoc} registration, alignment, and stitching of imaging data. Together, this architecture can enable flexible, customizable, and automated operation of a wide range of microscopy experiments.

\subsection{Operation System}

\begin{figure}
\includegraphics[width=\textwidth]{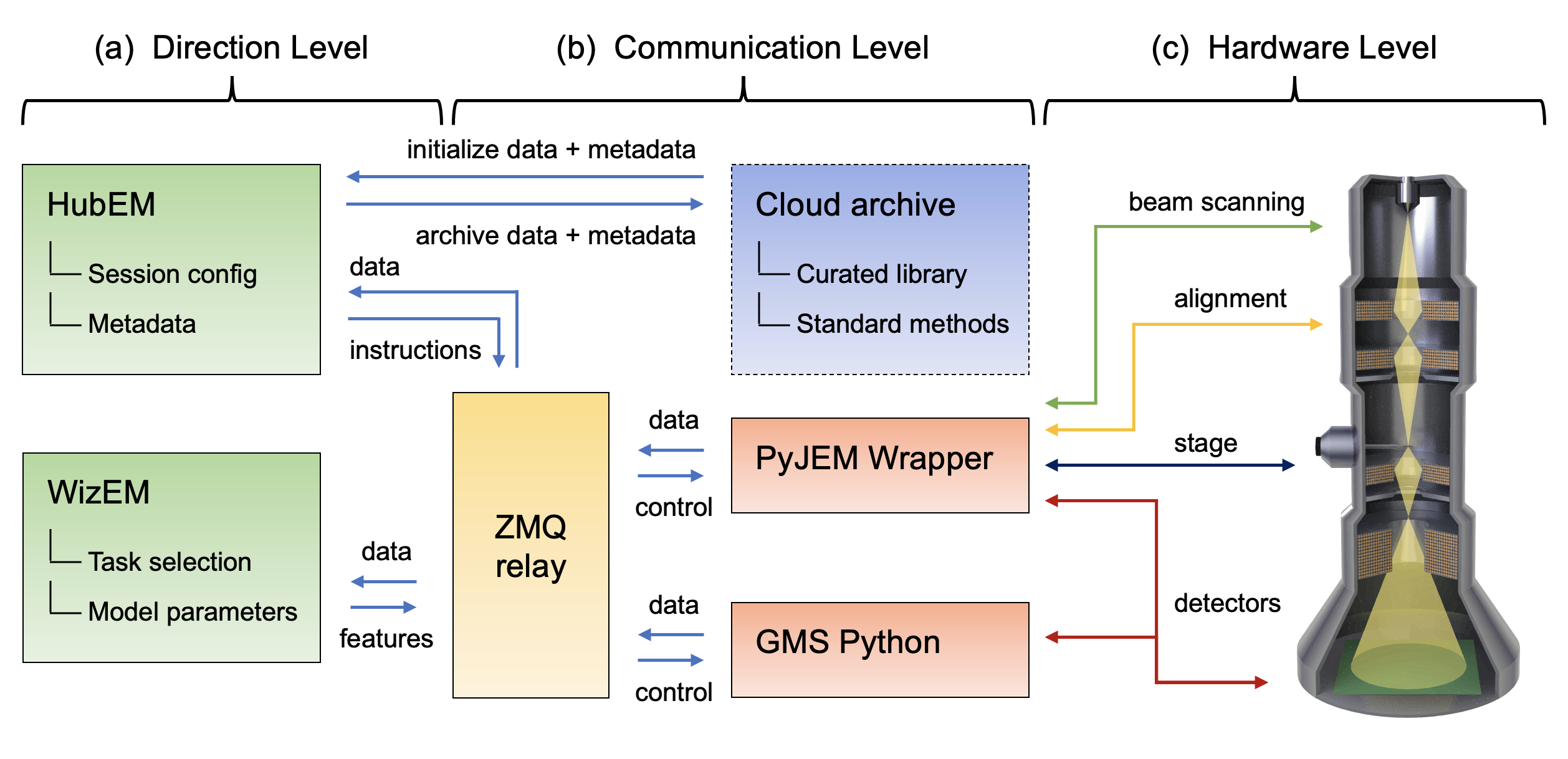}
\caption{Illustration of the operation system architecture, consisting of (a--c) Direction, Communication, and Hardware Levels, respectively. Arrows indicate the flow of signals between different hardware and software components at different levels of the architecture. \label{overview}}
\end{figure}

As shown in Figure \ref{overview}, the backbone of this platform is a distributed Operation System for acquiring image data in an open-loop fashion, analyzing that data via few-shot ML, and then optionally automatically deciding on the next steps of an experiment in a closed-loop fashion. The distributed nature of the system allows for analysis execution on a separate dedicated ML station, which is optimized for parallel processing, acquisition and control of various instruments in a remote lab, as well as remote visualization of the process from the office or home. This remote visualization stands in contrast to remote operating schemes, which can suffer from latency and communication drop outs that impact reliability.

The operation system consists of three levels: a \textit{Direction Level}, \textit{Communication Level}, and \textit{Hardware Level}. The Direction Level (Figure \ref{overview}a) includes two applications, HubEM and WizEM, designed for overall operation and few-shot ML analysis, respectively. These applications are the primary means for the end-user to interact with the microscope once a sample has been loaded and initial alignments have been performed. Each of these components is a separate process and may run on separate machines. HubEM is the main data acquisition application. It sends session configuration information to instrument controllers, receives data/metadata from them, and collates this information for a given experiment. HubEM passes instrument data to WizEM for few-shot ML analysis and receives analyzed data back for storage and real-time visualization. WizEM is a few-shot ML application featuring a web-based Python Flask GUI.\cite{Doty2021} It is used to classify and record the quantity and coordinates of user-defined features in images. The results of the analysis can be displayed to the user at the end of an open-loop acquisition or used as the basis for closed-loop decision making, as described in Section \ref{sec_control}.

Next, we consider the Communication Level shown Figure \ref{overview}b, which connects the end-user applications to low-level hardware commands. This level is intentionally designed to minimize the amount of direct user interaction with multiple hardware sub-systems, a process that can be slow and error-prone in more traditional microscope systems. Communications between various parts of the system are handled by a central messaging relay implemented in ZeroMQ (ZMQ), a socket-based messaging protocol that has been ported to many software languages and hardware platforms.\cite{Authors2021} The ZMQ publisher/subscriber model was chosen because it allows for asynchronous communication; that is, a component can publish a message and then continue with its work. For instance, HubEM can publish an image on a port subscribed to by WizEM. HubEM then continues its work of directing image acquisition, storage, and visualization, while periodically checking the WizEM port to which it subscribes. Concurrently, the WizEM code ``listens'' for any messages from HubEM via the ZMQ relay; these messages will contain the image to be processed by the few-shot script. WizEM analyzes the inbound data with the necessary parameters for few-shot analysis (as explained in Section \ref{sec_control} and Figure \ref{control}). These parameters are selected by the user via the WizEM GUI at the start of, or at decision points during, an experiment. The output from the few-shot analysis is the processed image, the coordinates of each classified feature, and summary statistics for display in HubEM. The WizEM code sends the analysis output back to the ZMQ relay, where it can be received by HubEM when resources are available. The HubEM application can be connected to Pacific Northwest National Laboratory's (PNNL) institutional data repository, known as DataHub.\cite{Laboratory2021} Data and methods, such as few-shot support sets, autoencoder selection, and model weights, can be initialized prior to an experiment and then uploaded at its conclusion.

The final, and lowest-level, component of the operation system is the Hardware Level, shown in Figure \ref{overview}c. This level has typically been the most challenging to implement, since direct low-level hardware controls are often unavailable or encoded in proprietary manufacturer formats. While many manufacturers have offered their own scripting languages,\cite{Mitchell2005a} these are usually inaccessible outside of siloed and limited application environments, which are incompatible with open Python or C++ / C\#-based programming languages. However, the recent release of APIs such as PyJEM\cite{PyJEM} and Gatan Microscopy Suite (GMS) Python\cite{Gatan2021a} has finally unlocked the ability to directly interface with most critical instrument operations, including beam control, alignment, stage positioning, and detectors. We have developed a wrapper for each of these APIs to define higher level controls that are then passed through the ZMQ relay. The Hardware Level is designed to be modular and can be extended through additional wrappers as new hardware is made accessible or additional components are installed. Together, the three levels of the Operation System provide distributed control of the microscope, linking it to rich automation and analysis applications via an asynchronous communications relay.

\clearpage

\subsection{Control System}\label{sec_control}

\begin{figure}
\includegraphics[width=\textwidth]{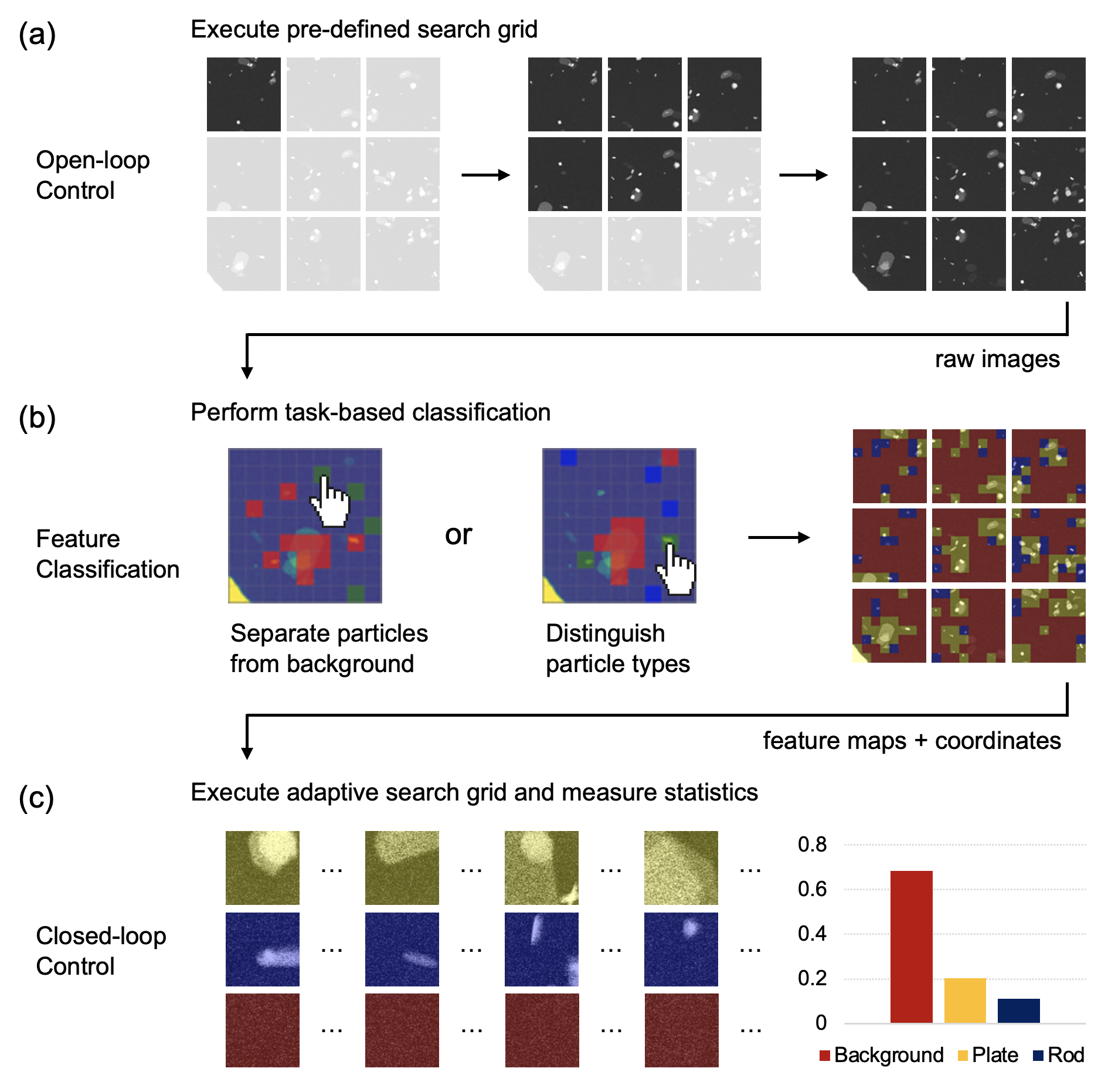}
\caption{Illustration of the control system architecture. (a) Open-loop Control generates search grid data that is passed through (b) Few-shot ML Feature Classification, informing optional (c) Closed-loop Control for complete automation.
\label{control}}
\end{figure}

 With the operation system in place, we can now implement various instrument control modes for specific experiments. As shown in Figure \ref{control}, the instrument can be run under \textit{Open-loop Control} or \textit{Closed-loop Control}, separated by a process of \textit{Feature Classification}. In Open-loop Control, the system executes a pre-defined search grid based on parameters provided by the user in the HubEM or downloaded from the DataHub institutional archive. The former approach may be used in everyday scenarios, when a user is unsure of the microstructural features contained within a sample, while the latter may be used in established large-scale screening campaigns of the same sample types or desired features. An advantage of this approach is that sampling methods can easily be standardized and shared among different instrument users or even among different laboratories.
 
 As data is acquired via Open-loop Control, it is passed to the WizEM application for feature classification shown in Figure \ref{control}b. The support set and model parameters for an analysis can dynamically adjusted in an interactive GUI, as described in Doty \textit{et al.},\citep{Doty2021} or be initialized from the cloud. In the typical application of the former process, the user selects one of the first few acquired frames containing microstructural features of interest. An adjustable grid is dynamically super-imposed on the image, and the image is separated at these grid lines into squares, called ``chips,'' of which a small fraction are subsequently assigned by the user into classes to define few-shot support sets. Using just 2--10 chips for each support set, the few-shot application runs a classification analysis on the current and subsequent images sent by HubEM. Each chip in each image is classified into one of the classes indicated in the support sets. The WizEM code incorporated into the Flask application sends the colorized segmented images, class coordinates, and summary statistics back to HubEM for real-time display to the user.
 
Once the user has defined features of interest for the few-shot analysis, it is possible to operate the instrument in a Closed-loop Control mode, shown in Figure \ref{control}c. In this mode, the initial search grid is executed to completion according to the user's specifications. The user then pre-selects feature types to target in a follow-up analysis (termed an ``adaptive search grid''). After the initial few-shot ML analysis is performed on each frame, the type and coordinates of each feature are identified and passed back to HubEM. The system then adjusts parameters such as stage coordinates, magnification, and detectors to automatically, adaptively sample desired feature types.

To illustrate a real-world example of instrument control, we consider the common use case of nanoparticle analysis. We have selected a sample of molybdenum trioxide (MoO$_3$) flakes, since it exhibits a range of particle sizes, orientations, and morphologies. MoO$_3$ is an important organic photovoltaic (OPVs) precursor,\cite{Gong2020a} has shown promise in preventing antimicrobial growth on surfaces,\cite{Zollfrank2012} and, when reduced to Mo, can provide corrosion resistance to austenitic stainless steels.\cite{Lyon2010} The TEM sample selected for this study has traditionally been utilized to calibrate diffraction rotation.\cite{Nakahara1992} For a diffraction rotation calibration, small, electron transparent platelets of varying dimension (100s nm to \textmu m) are evaporated onto a carbon film TEM grid.\cite{Doty2021}

As shown in Figure \ref{control}a, the user first acquires a pre-defined search grid within the HubEM application. This search grid is collected with specific image overlap parameters and knowledge of the stage movements to facilitate post-acquisition stitching, as will be discussed in Section \ref{sec_processing}. The observed distribution and orientation of the particles includes individual platelets lying both parallel and perpendicular to the primary beam (termed ``rod'' and ``plate,'' respectively), as well as plate clusters. Particle coordinates and type can then be measured automatically via few-shot ML analysis. To do this, the initial image frames in the open-loop acquisition are passed through the ZMQ relay for asynchronous analysis in the WizEM application. In this separate application, the user selects examples of the features of interest according to a desired task, which is an important advantage of the few-shot approach. As shown in Figure \ref{control}b, the few-shot model can, for example, be tuned to distinguish all particles from the background or to separate specific particle types (e.g., plates and rods) by selecting appropriate support sets. Importantly, this task can easily be adjusted on-the-fly or in \textit{post hoc} analysis as new information is acquired. Using this information, image segmentation, colorization, and statistical analysis of feature distributions is performed on subsequent data as the Open-loop Control proceeds. This information is passed back to the HubEM application, where it is presented dynamically to the user.

In the final mode of Closed-loop Control, the stage can drive to specific coordinates of identified particles, adjusting magnification or acquisition settings such as beam sampling or detector. This step is the most challenging part of the experiment, since it relies on precise recall of stage position and stability of instrument alignment. Here we propose a method for lower ($20-25$ kx) magnification Closed-loop Control, which is nonetheless valuable for many statistical analyses. At higher magnification, the stage is far more susceptible to mechanical imprecision and focus drift, which requires considerably more feedback in the control scheme. To better understand these challenges, we next consider the details of the acquisition and the important step of data processing for visualization and quantification.

\subsection{Data Processing System} \label{sec_processing}

\begin{figure}
\includegraphics[width=\textwidth]{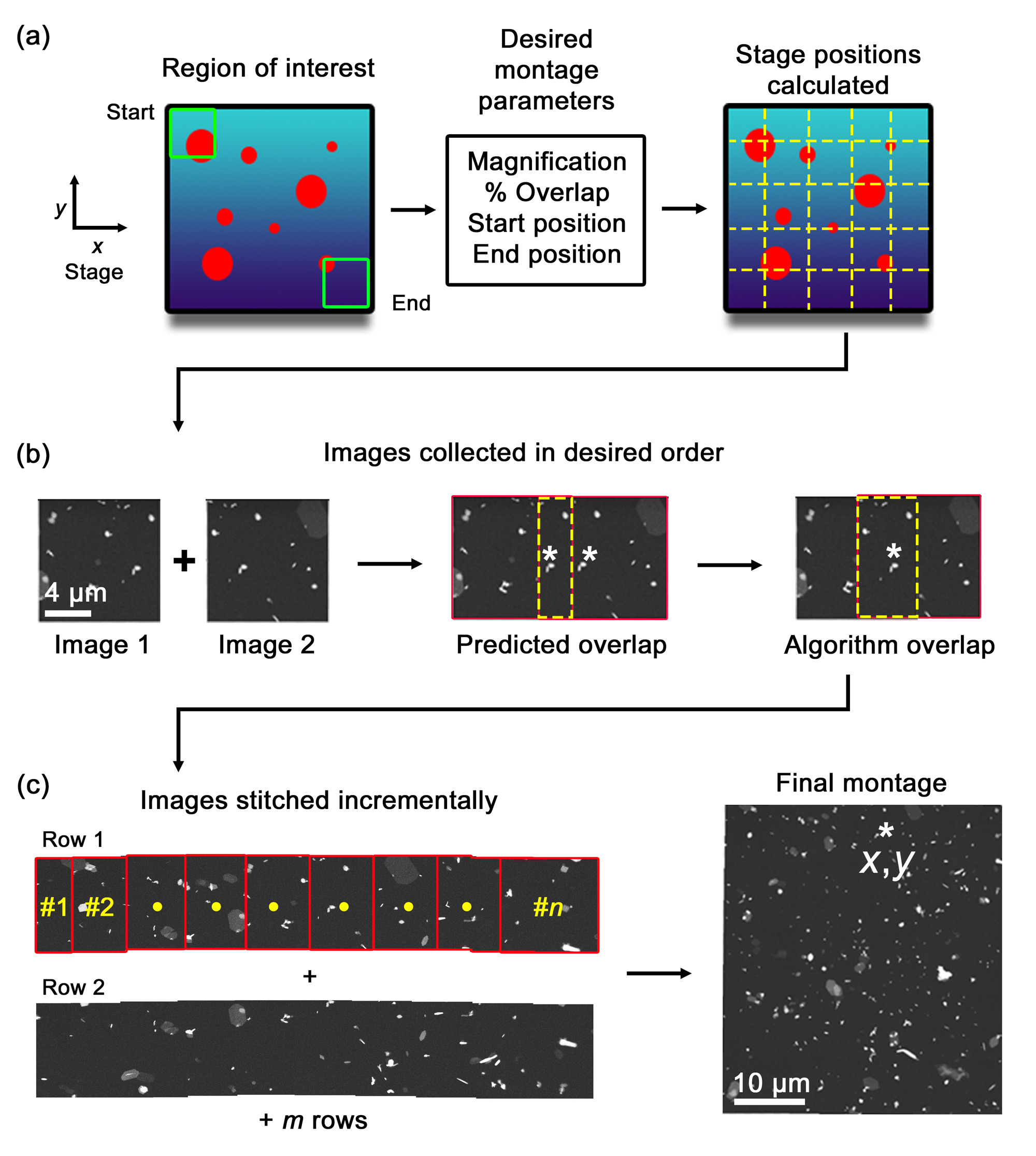}
\caption{MoO$_3$ data acquisition and processing. (a) Region of interest covered by montage and calculation of stage positions between a start and end point. (b) Calculation of stage positions and observed image overlap. (c) Incremental, row-by-row stitching of images conducted via cross-correlation to produce a final montage. \label{processing}}
\end{figure}

Alongside the Operation and Control Systems already described, we have developed a Data Processing System for large-area data collection, registration, and stitching of images. This processing is important to orient the user to the global position of local microstructural features and is needed for both closed-loop control and accurate statistical analysis. Building on the MoO$_3$ example discussed in Section \ref{sec_control}, we next consider the practical steps in the data acquisition process, as shown in Figure \ref{processing}. While this sample is ideal because it contains different particle morphologies and orientations, it is also challenging to analyze because of the sparsity of those particles (i.e., large fraction of empty carbon background). It is therefore necessary to perform lower magnification montaging in such a way that adjacent images are overlapped in both the $x$ and $y$ stage direction. First, the user selects a single ROI within the Cu TEM grid with no tears and a high density of particles, as shown schematically in Figure \ref{processing}a, with the closed red circles representing a desired feature on a support grid denoted by the blue background. This ROI is typically selected at lower magnification to increase the overall field of view (FOV), but may also be selected at higher magnification. Alternatively, fiducial markers, such as the corners of a finder grid, may be used to define the ROI. In either case, the $x$ and $y$ coordinates at the opposite corners of the ROI are defined as the collection Start and End positions, respectively. 

Upon selection of the desired ROI, the user is prompted to enter both the magnification and the desired percent overlap between consecutive images in the montage. Combined with the Start and End positions, the montaging algorithm calculates both the number of frames as well as the stage coordinates of each individual image to be collected. Depending on user's preference, the system can collect each image in a \textit{serpentine} or a \textit{sawtooth-raster} search pattern, the latter of which is commonly used in commercial acquisition systems. In the serpentine pattern, a search is conducted starting in the upper left corner and moving to the right until reaching the end of the row ($n$th frame). The search then moves down one row and back towards the left, repeating row-by-row until the $m$th row is reached and the montage is complete. In the sawtooth-raster, the search pattern also starts in the upper left, moving to the right until the end of the row is reached, just as in the serpentine pattern. At the end of the row, the direction of movement is reversed all the way back to the starting position before moving down to the next row, akin to the movement of a typewriter. From a montaging and image processing perspective there is little practical difference between these two methods, so the reduced travel time of the serpentine method is typically preferred. However, imprecision in stage movements (e.g., mechanical lash and flyback error) can lead to deviation in the precision of these approaches, particularly at higher magnification.

As shown in Figure \ref{processing}b, after selection of the ROI and start of the acquisition, the program collects the first image (Image 1), at which time the feature classification process in Figure \ref{control}b can be used to define the classification task. The acquisition is optionally paused while this step is conducted, and then a second image is acquired (Image 2); the software then utilizes the predicted overlap coordinates to perform an initial image alignment check. The montaging algorithm, described next, is then employed for further refinement of the relative displacement between the two images needed for feature alignment. We note that in Figure \ref{processing}b (Predicted overlap) the same particles (white asterisks) are observed in each image, but are not overlapped with one another. When further refinement of the montaging algorithm is applied (Algorithm overlap), the particles overlap (again noted by the white asterisk). Upon completion of the $n$ frames in Row 1, the acquisition proceeds until the $m$th row of data is collected and the End position is reached. While shown in Figure \ref{processing}c as complete rows, during operation each image is stitched incrementally to previous data collected in real-time. Once all stage positions have been imaged, a Final montage is calculated, at which time the user has the ability to manually or automatically select a region of interest (e.g., the particles denoted by the white asterisk) in order for the microscope to drive to the desired position and magnification.

When montaging is based solely on image capture, especially over large areas, there are many potential complications that can affect the final stitched montage. Depending on the STEM imaging conditions selected, beam drift can push the scattered diffraction discs closer to a given detector (e.g., strong diffraction onto a dark-field detector) that can skew imaging conditions from the first to the last image collected. As already mentioned, particle sparsity or clustering within the ROI can also present difficulties. For example, if the magnification is set too high, there may be regions within adjacent areas that have no significant contrast or features for registration. Such a situation might be encountered in large area particle analysis, as well as in grain distributions of uniform contrast. Understanding the stage motion is imperative in these cases, because the predicted image position can be utilized. Lastly, imprecise stage motion and image timing are important considerations. If the stage is moving during image capture, images can become blurred. In addition, if the area of interest is too large, sample height change can affect the image quality due to large defocus.

In light of these complications, we have evaluated image stitching approaches based on knowledge of the stage motion, as well as those solely based on image features. In principle, the simplest method is the former, in which prediction based on stage motion is used to calculated the overlap between two images directly. However, this method leads to artifacts in the stitched image due to a variety of practical factors related to high-throughput stage movement and image acquisition.\cite{Yin2020} For example, in some instances, motor hysteresis or stage lash causes a stage position to deviate from an issued command. An example of where the ``predicted overlap'' fails to accurately stitch adjacent images is shown in Figure \ref{processing}b. Image-by-image corrections must therefore be performed \textit{post hoc} using either manual or automated approaches. Manual stitching works surprisingly well for small numbers of images because the human eye is good at detecting patterns. However, this process is very time intensive, does not scale well to large montages, and cannot be automated within a program for automatic acquisition. To automate this process, the community has developed several standalone software packages,\cite{schneider_2012, rueden_2017, schindelin_2012} but these do not provide the user with immediate feedback while directly interfacing with the microscope. As part of the processing system, we have developed an image-based registration script to dynamically align and stitch images during an acquisition. At a high level, the algorithm functions by computing the cross correlation of adjacent images quickly using the Convolution Theorem as implemented in SciPy's signal processing library,\cite{2020SciPy-NMeth} and then identifies the peak of the cross correlation to find the correct displacement for maximum alignment. From this normalized cross correlation, the best alignment is determined from a the local maximum closest to predicted overlap, which is shown in the right portion of Figure \ref{processing}b, labeled ``Algorithm Overlap.'' This alignment process then repeats for every image as it is acquired to build up the overall montage. After processing the raw images, the same corrections can be applied to the few-shot classified montage, providing the user with a global survey of statistics on feature distributions in their sample.

\clearpage

\section{Conclusions}

We demonstrate the design of an automation system combining low-level instrument communication with closed-loop control based on few-shot ML. This system provides a practical abstraction of low-level hardware components from multiple manufacturers, which can be easily programmed through intuitive GUI applications by the end-user. It allows for microscope operation in both open-loop and closed-loop fashions, permitting task-based, high-throughput statistical analysis, in contrast to more traditional automation approaches that are more labor-intensive and less flexible.

Future developments will continue to improve and extend the functionality of the automation system. In particular, reliable stage motion is a crucial part of the data acquisition process; better understanding of stages and improvements in hardware will help address current shortcomings, such as atomic-resolution montaging. Detailed corrections for focusing and beam alignment will also become important during high-magnification acquisitions. The integration of additional imaging modalities, such as diffraction and spectroscopy, is possible and will greatly extend the utility of the system. Future designs may also integrate richer physics-based ML models or additional processing steps to enhance feature detection and improve the system control loop. This research presents a model for more integrated and scalable control of electron microscopy. Moving forward, increasingly automated, and eventually autonomous approaches will enable richer and more standardizable experimentation, helping to transform the process of discovery across all scientific domains.

\clearpage

\section{Acknowledgements}

We thank Dr. Elizabeth Kautz for reviewing the manuscript. This research was supported by the I3T Commercialization Laboratory Directed Research and Development (LDRD) program at Pacific Northwest National Laboratory (PNNL). PNNL is a multiprogram national laboratory operated for the U.S. Department of Energy (DOE) by Battelle Memorial Institute under Contract No. DE-AC05-76RL0-1830. The few-shot ML code development was also supported by the Chemical Dynamics Initiative (CDi) LDRD program, with some initial code development performed under the Nuclear Process Science Initiative (NPSI) LDRD program. The system was constructed in the Radiological Microscopy Suite (RMS), located in the Radiochemical Processing Laboratory (RPL) at PNNL. Some sample preparation was performed at the Environmental Molecular Sciences Laboratory (EMSL), a national scientific user facility sponsored by the Department of Energy's Office of Biological and Environmental Research and located at PNNL.

\section{Competing Interests Statement}

The authors declare no competing interests.

\section{Data Availability Statement}

The raw and processed MoO$_3$ frames shown are available on FigShare at \url{https://doi.org/10.6084/m9.figshare.14850102.v2} Additional details on the scripting and hardware configuration are available from the authors upon reasonable request.

\clearpage

\bibliography{refs_steven, refs_oostrom, refs_fiedler}

\begin{thebibliography}{10}

\bibitem{Daston2011}
Lorraine Daston and Lunbeck Elizabeth, editors.
\newblock {\em {Histories of Scientific Observation}}.
\newblock University of Chicago Press, Chicago and London, 2011.

\bibitem{Zhang2019}
Chao Zhang, Konstantin~L. Firestein, Joseph F.~S. Fernando, Dumindu
  Siriwardena, Joel~E. Treifeldt, and Dmitri Golberg.
\newblock {Recent Progress of In Situ Transmission Electron Microscopy for
  Energy Materials}.
\newblock {\em Adv. Mater.}, 1904094:1904094, sep 2019.
\newblock URL:
  \url{https://onlinelibrary.wiley.com/doi/abs/10.1002/adma.201904094}, \href
  {https://doi.org/10.1002/adma.201904094} {\path{doi:10.1002/adma.201904094}}.

\bibitem{VanBenthem2007}
Klaus {Van Benthem} and Stephen~J Pennycook.
\newblock {Aberration-corrected Scanning Transmission Electron Microscopy for
  Atomic-scale Characterization of Semiconductor Devices}.
\newblock {\em ECS Trans.}, 11(3):225--231, dec 2007.
\newblock URL: \url{https://iopscience.iop.org/article/10.1149/1.2779062},
  \href {https://doi.org/10.1149/1.2779062} {\path{doi:10.1149/1.2779062}}.

\bibitem{Shen2018}
Peter~S. Shen.
\newblock {The 2017 Nobel Prize in Chemistry: cryo-EM comes of age}.
\newblock {\em Anal. Bioanal. Chem.}, 410(8):2053--2057, 2018.
\newblock \href {https://doi.org/10.1007/s00216-018-0899-8}
  {\path{doi:10.1007/s00216-018-0899-8}}.

\bibitem{Frank2017}
Joachim Frank.
\newblock {Advances in the field of single-particle cryo-electron microscopy
  over the last decade}.
\newblock {\em Nat. Protoc.}, 12(2):209--212, 2017.
\newblock \href {https://doi.org/10.1038/nprot.2017.004}
  {\path{doi:10.1038/nprot.2017.004}}.

\bibitem{Gupta2018}
Shivam Gupta, Arpan~Kumar Kar, Abdullah Baabdullah, and Wassan~A.A.
  Al-Khowaiter.
\newblock {Big data with cognitive computing: A review for the future}.
\newblock {\em Int. J. Inf. Manage.}, 42(June):78--89, oct 2018.
\newblock URL: \url{https://doi.org/10.1016/j.ijinfomgt.2018.06.005
  https://linkinghub.elsevier.com/retrieve/pii/S0268401218304110}, \href
  {https://doi.org/10.1016/j.ijinfomgt.2018.06.005}
  {\path{doi:10.1016/j.ijinfomgt.2018.06.005}}.

\bibitem{Baraniuk2011}
Richard~G. Baraniuk.
\newblock {More Is Less: Signal Processing and the Data Deluge}.
\newblock {\em Science (80-. ).}, 331(6018):717--719, feb 2011.
\newblock URL:
  \url{https://www.sciencemag.org/lookup/doi/10.1126/science.1197448}, \href
  {https://doi.org/10.1126/science.1197448}
  {\path{doi:10.1126/science.1197448}}.

\bibitem{Stach2021}
Eric Stach, Brian DeCost, A~Gilad Kusne, Jason Hattrick-Simpers, Keith~A Brown,
  Kristofer~G Reyes, Joshua Schrier, Simon Billinge, Tonio Buonassisi, Ian
  Foster, Carla~P Gomes, John~M Gregoire, Apurva Mehta, Joseph Montoya, Elsa
  Olivetti, Chiwoo Park, Eli Rotenberg, Semion~K Saikin, Sylvia Smullin,
  Valentin Stanev, and Benji Maruyama.
\newblock {Autonomous experimentation systems for materials development: A
  community perspective}.
\newblock {\em Matter}, pages 1--25, jul 2021.
\newblock URL: \url{https://doi.org/10.1016/j.matt.2021.06.036
  https://linkinghub.elsevier.com/retrieve/pii/S2590238521003064}, \href
  {https://doi.org/10.1016/j.matt.2021.06.036}
  {\path{doi:10.1016/j.matt.2021.06.036}}.

\bibitem{Noack2021}
Marcus~M Noack, Petrus~H. Zwart, Daniela~M. Ushizima, Masafumi Fukuto, Kevin~G.
  Yager, Katherine~C. Elbert, Christopher~B. Murray, Aaron Stein, Gregory~S
  Doerk, Esther H.~R. Tsai, Ruipeng Li, Guillaume Freychet, Mikhail Zhernenkov,
  Hoi-Ying~N. Holman, Steven Lee, Liang Chen, Eli Rotenberg, Tobias Weber,
  Yannick~Le Goc, Martin Boehm, Paul Steffens, Paolo Mutti, and James~A
  Sethian.
\newblock {Gaussian processes for autonomous data acquisition at large-scale
  synchrotron and neutron facilities}.
\newblock {\em Nat. Rev. Phys.}, 0123456789, jul 2021.
\newblock URL: \url{http://dx.doi.org/10.1038/s42254-021-00345-y
  http://www.nature.com/articles/s42254-021-00345-y}, \href
  {https://doi.org/10.1038/s42254-021-00345-y}
  {\path{doi:10.1038/s42254-021-00345-y}}.

\bibitem{Brown2020}
Keith~A. Brown, Sarah Brittman, Nicol{\`{o}} Maccaferri, Deep Jariwala, and
  Umberto Celano.
\newblock {Machine Learning in Nanoscience: Big Data at Small Scales}.
\newblock {\em Nano Lett.}, 20(1):2--10, jan 2020.
\newblock URL: \url{https://pubs.acs.org/doi/10.1021/acs.nanolett.9b04090},
  \href {https://doi.org/10.1021/acs.nanolett.9b04090}
  {\path{doi:10.1021/acs.nanolett.9b04090}}.

\bibitem{Vasudevan2019}
Rama~K. Vasudevan, Kamal Choudhary, Apurva Mehta, Ryan Smith, Gilad Kusne,
  Francesca Tavazza, Lukas Vlcek, Maxim Ziatdinov, Sergei~V. Kalinin, and Jason
  Hattrick-Simpers.
\newblock {Materials science in the artificial intelligence age:
  high-throughput library generation, machine learning, and a pathway from
  correlations to the underpinning physics}.
\newblock {\em MRS Commun.}, 9(03):821--838, sep 2019.
\newblock URL:
  \url{https://www.cambridge.org/core/product/identifier/S2159685919000958/type/journal{\_}article},
  \href {https://doi.org/10.1557/mrc.2019.95} {\path{doi:10.1557/mrc.2019.95}}.

\bibitem{King2004}
Ross~D. King, Kenneth~E. Whelan, Ffion~M. Jones, Philip G.~K. Reiser,
  Christopher~H. Bryant, Stephen~H. Muggleton, Douglas~B. Kell, and Stephen~G.
  Oliver.
\newblock {Functional genomic hypothesis generation and experimentation by a
  robot scientist}.
\newblock {\em Nature}, 427(6971):247--252, jan 2004.
\newblock URL: \url{http://www.nature.com/articles/nature02236}, \href
  {https://doi.org/10.1038/nature02236} {\path{doi:10.1038/nature02236}}.

\bibitem{Xu2020}
Yanheng Xu, Rulin Liu, Jiagen Li, Yao Xu, and Xi~Zhu.
\newblock {The Blockchain Integrated Automatic Experiment Platform (BiaeP)}.
\newblock {\em J. Phys. Chem. Lett.}, 11(23):9995--10000, dec 2020.
\newblock URL: \url{https://pubs.acs.org/doi/10.1021/acs.jpclett.0c02840},
  \href {https://doi.org/10.1021/acs.jpclett.0c02840}
  {\path{doi:10.1021/acs.jpclett.0c02840}}.

\bibitem{Shields2021}
Benjamin~J. Shields, Jason Stevens, Jun Li, Marvin Parasram, Farhan Damani,
  Jesus~I.Martinez Alvarado, Jacob~M. Janey, Ryan~P. Adams, and Abigail~G.
  Doyle.
\newblock {Bayesian reaction optimization as a tool for chemical synthesis}.
\newblock {\em Nature}, 590(7844):89--96, 2021.
\newblock URL: \url{http://dx.doi.org/10.1038/s41586-021-03213-y}, \href
  {https://doi.org/10.1038/s41586-021-03213-y}
  {\path{doi:10.1038/s41586-021-03213-y}}.

\bibitem{Higgins2020}
Kate Higgins, Sai~Mani Valleti, Maxim Ziatdinov, Sergei~V. Kalinin, and Mahshid
  Ahmadi.
\newblock {Chemical Robotics Enabled Exploration of Stability in Multicomponent
  Lead Halide Perovskites via Machine Learning}.
\newblock {\em ACS Energy Lett.}, 5(11):3426--3436, nov 2020.
\newblock URL: \url{https://pubs.acs.org/doi/10.1021/acsenergylett.0c01749},
  \href {https://doi.org/10.1021/acsenergylett.0c01749}
  {\path{doi:10.1021/acsenergylett.0c01749}}.

\bibitem{Hase2019}
Florian H{\"{a}}se, Lo{\"{i}}c~M. Roch, and Al{\'{a}}n Aspuru-Guzik.
\newblock {Next-Generation Experimentation with Self-Driving Laboratories}.
\newblock {\em Trends Chem.}, 1(3):282--291, jun 2019.
\newblock URL:
  \url{https://linkinghub.elsevier.com/retrieve/pii/S258959741930019X}, \href
  {https://doi.org/10.1016/j.trechm.2019.02.007}
  {\path{doi:10.1016/j.trechm.2019.02.007}}.

\bibitem{Takeuchi2005}
Ichiro Takeuchi, Jochen Lauterbach, and Michael~J. Fasolka.
\newblock {Combinatorial materials synthesis}.
\newblock {\em Mater. Today}, 8(10):18--26, oct 2005.
\newblock URL: \url{http://dx.doi.org/10.1016/S1369-7021(05)71121-4
  https://linkinghub.elsevier.com/retrieve/pii/S1369702105711214}, \href
  {https://doi.org/10.1016/S1369-7021(05)71121-4}
  {\path{doi:10.1016/S1369-7021(05)71121-4}}.

\bibitem{Arzt2005}
Steffi Arzt, Antonia Beteva, Florent Cipriani, Solange Delageniere, Franck
  Felisaz, Gabriele F{\"{o}}rstner, Elspeth Gordon, Ludovic Launer, Bernard
  Lavault, Gordon Leonard, Trevor Mairs, Andrew McCarthy, Joanne McCarthy, Sean
  McSweeney, Jens Meyer, Edward Mitchell, Stephanie Monaco, Didier Nurizzo,
  Raimond Ravelli, Vicente Rey, William Shepard, Darren Spruce, Olof Svensson,
  and Pascal Theveneau.
\newblock {Automation of macromolecular crystallography beamlines}.
\newblock {\em Prog. Biophys. Mol. Biol.}, 89(2):124--152, oct 2005.
\newblock URL:
  \url{https://linkinghub.elsevier.com/retrieve/pii/S0079610704001233}, \href
  {https://doi.org/10.1016/j.pbiomolbio.2004.09.003}
  {\path{doi:10.1016/j.pbiomolbio.2004.09.003}}.

\bibitem{Abola2000}
Enrique Abola, Peter Kuhn, Thomas Earnest, and Raymond~C Stevens.
\newblock {Automation of X-ray crystallography}.
\newblock {\em Nature}, (november):973--977, 2000.
\newblock \href {https://doi.org/10.1038/80754} {\path{doi:10.1038/80754}}.

\bibitem{Nogales2015}
Eva Nogales.
\newblock {The development of cryo-EM into a mainstream structural biology
  technique}.
\newblock {\em Nat. Methods}, 13(1):24--27, jan 2016.
\newblock URL: \url{http://www.nature.com/doifinder/10.1038/nmeth.3694
  http://www.nature.com/articles/nmeth.3694}, \href
  {https://doi.org/10.1038/nmeth.3694} {\path{doi:10.1038/nmeth.3694}}.

\bibitem{Carragher2000}
Bridget Carragher, Nick Kisseberth, David Kriegman, Ronald~A. Milligan,
  Clinton~S. Potter, James Pulokas, and Amy Reilein.
\newblock {Leginon: An Automated System for Acquisition of Images from Vitreous
  Ice Specimens}.
\newblock {\em J. Struct. Biol.}, 132(1):33--45, oct 2000.
\newblock URL:
  \url{https://linkinghub.elsevier.com/retrieve/pii/S1047847700943144}, \href
  {https://doi.org/10.1006/jsbi.2000.4314} {\path{doi:10.1006/jsbi.2000.4314}}.

\bibitem{Hattar2021}
Khalid Hattar and Katherine~L. Jungjohann.
\newblock {Possibility of an integrated transmission electron microscope:
  enabling complex in-situ experiments}.
\newblock {\em J. Mater. Sci.}, 56(9):5309--5320, 2021.
\newblock \href {https://doi.org/10.1007/s10853-020-05598-z}
  {\path{doi:10.1007/s10853-020-05598-z}}.

\bibitem{Ede2020}
Jeffrey~M. Ede.
\newblock {Deep learning in electron microscopy}.
\newblock {\em Mach. Learn. Sci. Technol.}, 2(1):011004, mar 2021.
\newblock URL: \url{http://arxiv.org/abs/2009.08328
  https://iopscience.iop.org/article/10.1088/2632-2153/abd614}, \href
  {http://arxiv.org/abs/2009.08328} {\path{arXiv:2009.08328}}, \href
  {https://doi.org/10.1088/2632-2153/abd614}
  {\path{doi:10.1088/2632-2153/abd614}}.

\bibitem{Schorb2019}
Martin Schorb, Isabella Haberbosch, Wim~J.H. Hagen, Yannick Schwab, and
  David~N. Mastronarde.
\newblock {Software tools for automated transmission electron microscopy}.
\newblock {\em Nat. Methods}, 16(6):471--477, 2019.
\newblock URL: \url{http://dx.doi.org/10.1038/s41592-019-0396-9}, \href
  {https://doi.org/10.1038/s41592-019-0396-9}
  {\path{doi:10.1038/s41592-019-0396-9}}.

\bibitem{Kalinin2015}
Sergei~V. Kalinin, Bobby~G. Sumpter, and Richard~K. Archibald.
\newblock {Big–deep–smart data in imaging for guiding materials design}.
\newblock {\em Nat. Mater.}, 14(10):973--980, sep 2015.
\newblock URL: \url{http://dx.doi.org/10.1038/nmat4395
  http://www.nature.com/doifinder/10.1038/nmat4395
  http://linkinghub.elsevier.com/retrieve/pii/S0261306906000628}, \href
  {https://doi.org/10.1038/nmat4395} {\path{doi:10.1038/nmat4395}}.

\bibitem{Spurgeon2020c}
Steven~R Spurgeon, Colin Ophus, Lewys Jones, Amanda Petford-Long, Sergei~V
  Kalinin, Matthew~J Olszta, Rafal~E. Dunin-Borkowski, Norman Salmon, Khalid
  Hattar, Wei-chang~D Yang, Renu Sharma, Yingge Du, Ann Chiaramonti, Haimei
  Zheng, Edgar~C Buck, Libor Kovarik, R~Lee Penn, Dongsheng Li, Xin Zhang,
  Mitsuhiro Murayama, and Mitra~L. Taheri.
\newblock {Towards data-driven next-generation transmission electron
  microscopy}.
\newblock {\em Nat. Mater.}, 20(3):274--279, mar 2021.
\newblock URL: \url{http://dx.doi.org/10.1038/s41563-020-00833-z
  http://www.nature.com/articles/s41563-020-00833-z}, \href
  {https://doi.org/10.1038/s41563-020-00833-z}
  {\path{doi:10.1038/s41563-020-00833-z}}.

\bibitem{Kalidindi2015}
Surya~R. Kalidindi and Marc {De Graef}.
\newblock {Materials Data Science: Current Status and Future Outlook}.
\newblock {\em Annu. Rev. Mater. Res.}, 45(1):171--193, jul 2015.
\newblock URL:
  \url{http://www.annualreviews.org/doi/10.1146/annurev-matsci-070214-020844},
  \href {https://doi.org/10.1146/annurev-matsci-070214-020844}
  {\path{doi:10.1146/annurev-matsci-070214-020844}}.

\bibitem{MaiaChagas2018}
Andr{\'{e}} {Maia Chagas}.
\newblock {Haves and have nots must find a better way: The case for open
  scientific hardware}.
\newblock {\em PLOS Biol.}, 16(9):e3000014, sep 2018.
\newblock URL: \url{https://dx.plos.org/10.1371/journal.pbio.3000014}, \href
  {https://doi.org/10.1371/journal.pbio.3000014}
  {\path{doi:10.1371/journal.pbio.3000014}}.

\bibitem{Mastronarde2003a}
David~N. Mastronarde.
\newblock {SerialEM: A Program for Automated Tilt Series Acquisition on Tecnai
  Microscopes Using Prediction of Specimen Position}.
\newblock {\em Microsc. Microanal.}, 9(S02):1182--1183, aug 2003.
\newblock URL:
  \url{https://www.cambridge.org/core/product/identifier/S1431927603445911/type/journal{\_}article},
  \href {https://doi.org/10.1017/S1431927603445911}
  {\path{doi:10.1017/S1431927603445911}}.

\bibitem{Yin2020}
Wenjing Yin, Derrick Brittain, Jay Borseth, Marie~E. Scott, Derric Williams,
  Jedediah Perkins, Christopher~S. Own, Matthew Murfitt, Russel~M. Torres,
  Daniel Kapner, Gayathri Mahalingam, Adam Bleckert, Daniel Castelli, David
  Reid, Wei-Chung~Allen Lee, Brett~J. Graham, Marc Takeno, Daniel~J. Bumbarger,
  Colin Farrell, R.~Clay Reid, and Nuno~Macarico da~Costa.
\newblock {A petascale automated imaging pipeline for mapping neuronal circuits
  with high-throughput transmission electron microscopy}.
\newblock {\em Nat. Commun.}, 11(1):4949, dec 2020.
\newblock URL: \url{http://www.nature.com/articles/s41467-020-18659-3}, \href
  {https://doi.org/10.1038/s41467-020-18659-3}
  {\path{doi:10.1038/s41467-020-18659-3}}.

\bibitem{Coudray2011}
Nicolas Coudray, Gilles Hermann, Daniel Caujolle-Bert, Argyro Karathanou,
  Fran{\c{c}}oise Erne-Brand, Jean~Luc Buessler, Pamela Daum, Juergen~M.
  Plitzko, Mohamed Chami, Urs Mueller, Hubert Kihl, Jean~Philippe Urban,
  Andreas Engel, and Herv{\'{e}}~W. R{\'{e}}migy.
\newblock {Automated screening of 2D crystallization trials using transmission
  electron microscopy: A high-throughput tool-chain for sample preparation and
  microscopic analysis}.
\newblock {\em J. Struct. Biol.}, 173(2):365--374, 2011.
\newblock URL: \url{http://dx.doi.org/10.1016/j.jsb.2010.09.019}, \href
  {https://doi.org/10.1016/j.jsb.2010.09.019}
  {\path{doi:10.1016/j.jsb.2010.09.019}}.

\bibitem{Martin-Isla2020}
Carlos Martin-Isla, Victor~M. Campello, Cristian Izquierdo, Zahra
  Raisi-Estabragh, Bettina Bae{\ss}ler, Steffen~E. Petersen, and Karim Lekadir.
\newblock {Image-Based Cardiac Diagnosis With Machine Learning: A Review}.
\newblock {\em Front. Cardiovasc. Med.}, 7(January):1--19, jan 2020.
\newblock URL:
  \url{https://www.frontiersin.org/article/10.3389/fcvm.2020.00001/full}, \href
  {https://doi.org/10.3389/fcvm.2020.00001}
  {\path{doi:10.3389/fcvm.2020.00001}}.

\bibitem{Gurcan2009}
M.N. Gurcan, L.E. Boucheron, A.~Can, A.~Madabhushi, N.M. Rajpoot, and B.~Yener.
\newblock {Histopathological Image Analysis: A Review}.
\newblock {\em IEEE Rev. Biomed. Eng.}, 2:147--171, 2009.
\newblock URL: \url{http://ieeexplore.ieee.org/document/5299287/}, \href
  {https://doi.org/10.1109/RBME.2009.2034865}
  {\path{doi:10.1109/RBME.2009.2034865}}.

\bibitem{Liu2016a}
Jinxin Liu, Hongchang Li, Lei Zhang, Matthew Rames, Meng Zhang, Yadong Yu,
  Bo~Peng, C{\'{e}}sar~D{\'{i}}az Celis, April Xu, Qin Zou, Xu~Yang, Xuefeng
  Chen, and Gang Ren.
\newblock {Fully Mechanically Controlled Automated Electron Microscopic
  Tomography}.
\newblock {\em Sci. Rep.}, 6(June):1--12, 2016.
\newblock URL: \url{http://dx.doi.org/10.1038/srep29231}, \href
  {https://doi.org/10.1038/srep29231} {\path{doi:10.1038/srep29231}}.

\bibitem{Rauch2014}
E.~F. Rauch and M.~V{\'{e}}ron.
\newblock {Automated crystal orientation and phase mapping in TEM}.
\newblock {\em Mater. Charact.}, 98:1--9, dec 2014.
\newblock URL:
  \url{https://linkinghub.elsevier.com/retrieve/pii/S1044580314002514}, \href
  {https://doi.org/10.1016/j.matchar.2014.08.010}
  {\path{doi:10.1016/j.matchar.2014.08.010}}.

\bibitem{Strauss2013}
Michael Strauss and Mark Williamson.
\newblock {Automated transmission electron microscopy for defect review and
  metrology of Si devices}.
\newblock In {\em ASMC 2013 SEMI Adv. Semicond. Manuf. Conf.}, pages 366--370.
  IEEE, may 2013.
\newblock URL: \url{http://ieeexplore.ieee.org/document/6552761/}, \href
  {https://doi.org/10.1109/ASMC.2013.6552761}
  {\path{doi:10.1109/ASMC.2013.6552761}}.

\bibitem{Uusimaeki2019}
Toni Uusimaeki, Thorsten Wagner, Hans~Gerd Lipinski, and Ralf Kaegi.
\newblock {AutoEM: a software for automated acquisition and analysis of
  nanoparticles}.
\newblock {\em J. Nanoparticle Res.}, 21(6), 2019.
\newblock \href {https://doi.org/10.1007/s11051-019-4555-9}
  {\path{doi:10.1007/s11051-019-4555-9}}.

\bibitem{House2017}
Stephen~D. House, Yuxiang Chen, Rongchao Jin, and Judith~C. Yang.
\newblock {High-throughput, semi-automated quantitative STEM mass measurement
  of supported metal nanoparticles using a conventional TEM/STEM}.
\newblock {\em Ultramicroscopy}, 182:145--155, nov 2017.
\newblock URL:
  \url{https://linkinghub.elsevier.com/retrieve/pii/S0304399117301559}, \href
  {https://doi.org/10.1016/j.ultramic.2017.07.004}
  {\path{doi:10.1016/j.ultramic.2017.07.004}}.

\bibitem{Kalinin2021}
Sergei~V. Kalinin, Maxim Ziatdinov, Jacob Hinkle, Stephen Jesse, Ayana Ghosh,
  Kyle~P. Kelley, Andrew~R. Lupini, Bobby~G. Sumpter, and Rama~K. Vasudevan.
\newblock {Automated and Autonomous Experiments in Electron and Scanning Probe
  Microscopy}.
\newblock {\em ACS Nano}, 15(8):12604--12627, aug 2021.
\newblock URL: \url{https://pubs.acs.org/doi/10.1021/acsnano.1c02104}, \href
  {http://arxiv.org/abs/2103.12165} {\path{arXiv:2103.12165}}, \href
  {https://doi.org/10.1021/acsnano.1c02104}
  {\path{doi:10.1021/acsnano.1c02104}}.

\bibitem{PyJEM}
PyJEM.
\newblock {PyJEM}.
\newblock URL: \url{https://github.com/PyJEM/PyJEM}.

\bibitem{Meyer2019a}
Chris Meyer, Niklas Dellby, Jordan~A. Hachtel, Tracy Lovejoy, Andreas
  Mittelberger, and Ondrej Krivanek.
\newblock {Nion Swift: Open Source Image Processing Software for Instrument
  Control, Data Acquisition, Organization, Visualization, and Analysis Using
  Python.}
\newblock {\em Microsc. Microanal.}, 25(S2):122--123, aug 2019.
\newblock URL:
  \url{https://www.cambridge.org/core/product/identifier/S143192761900134X/type/journal{\_}article},
  \href {https://doi.org/10.1017/s143192761900134x}
  {\path{doi:10.1017/s143192761900134x}}.

\bibitem{Ziatdinov2021}
Maxim Ziatdinov, Ayana Ghosh, Tommy Wong, and Sergei~V. Kalinin.
\newblock {AtomAI: A Deep Learning Framework for Analysis of Image and
  Spectroscopy Data in (Scanning) Transmission Electron Microscopy and Beyond}.
\newblock pages 1--25, may 2021.
\newblock URL: \url{http://arxiv.org/abs/2105.07485}, \href
  {http://arxiv.org/abs/2105.07485} {\path{arXiv:2105.07485}}.

\bibitem{Tejada2011}
Arturo Tejada, Arnold~J. den Dekker, and Wouter {Van den Broek}.
\newblock {Introducing measure-by-wire, the systematic use of systems and
  control theory in transmission electron microscopy}.
\newblock {\em Ultramicroscopy}, 111(11):1581--1591, nov 2011.
\newblock URL: \url{http://dx.doi.org/10.1016/j.ultramic.2011.08.011
  https://linkinghub.elsevier.com/retrieve/pii/S0304399111002051}, \href
  {https://doi.org/10.1016/j.ultramic.2011.08.011}
  {\path{doi:10.1016/j.ultramic.2011.08.011}}.

\bibitem{Curtarolo2013}
Stefano Curtarolo, Gus L.~W. Hart, Marco~Buongiorno Nardelli, Natalio Mingo,
  Stefano Sanvito, and Ohad Levy.
\newblock {The high-throughput highway to computational materials design}.
\newblock {\em Nat. Mater.}, 12(3):191--201, 2013.
\newblock URL: \url{http://www.nature.com/doifinder/10.1038/nmat3568}, \href
  {https://doi.org/10.1038/nmat3568} {\path{doi:10.1038/nmat3568}}.

\bibitem{Taheri2016}
Mitra~L. Taheri, Eric~A. Stach, Ilke Arslan, P.~A. Crozier, Bernd~C. Kabius,
  Thomas LaGrange, Andrew~M. Minor, Seiji Takeda, Mihaela Tanase, Jakob~B.
  Wagner, and Renu Sharma.
\newblock {Current status and future directions for in situ transmission
  electron microscopy}.
\newblock {\em Ultramicroscopy}, 170:86--95, 2016.
\newblock URL: \url{http://dx.doi.org/10.1016/j.ultramic.2016.08.007}, \href
  {https://doi.org/10.1016/j.ultramic.2016.08.007}
  {\path{doi:10.1016/j.ultramic.2016.08.007}}.

\bibitem{Voyles2016}
Paul~M. Voyles.
\newblock {Informatics and data science in materials microscopy}.
\newblock {\em Curr. Opin. Solid State Mater. Sci.}, 21(3):141--158, jun 2017.
\newblock URL: \url{http://dx.doi.org/10.1016/j.cossms.2016.10.001
  http://linkinghub.elsevier.com/retrieve/pii/S135902861630095X
  https://linkinghub.elsevier.com/retrieve/pii/S135902861630095X}, \href
  {https://doi.org/10.1016/j.cossms.2016.10.001}
  {\path{doi:10.1016/j.cossms.2016.10.001}}.

\bibitem{Roccapriore2021}
Kevin~M. Roccapriore, Sergei~V. Kalinin, and Maxim Ziatdinov.
\newblock {Physics discovery in nanoplasmonic systems via autonomous
  experiments in Scanning Transmission Electron Microscopy}.
\newblock aug 2021.
\newblock URL: \url{http://arxiv.org/abs/2108.03290}, \href
  {http://arxiv.org/abs/2108.03290} {\path{arXiv:2108.03290}}.

\bibitem{Belianinov2015}
Alex Belianinov, Rama Vasudevan, Evgheni Strelcov, Chad Steed, Sang~Mo Yang,
  Alexander Tselev, Stephen Jesse, Michael Biegalski, Galen Shipman,
  Christopher Symons, Albina Borisevich, Rick Archibald, and Sergei Kalinin.
\newblock {Big data and deep data in scanning and electron microscopies:
  deriving functionality from multidimensional data sets}.
\newblock {\em Adv. Struct. Chem. Imaging}, 1(1):6, dec 2015.
\newblock URL: \url{http://www.ascimaging.com/content/1/1/6}, \href
  {https://doi.org/10.1186/s40679-015-0006-6}
  {\path{doi:10.1186/s40679-015-0006-6}}.

\bibitem{Madsen2018a}
Jacob Madsen, Pei Liu, Jens Kling, Jakob~Birkedal Wagner, Thomas~Willum Hansen,
  Ole Winther, and Jakob Schi{\o}tz.
\newblock {A Deep Learning Approach to Identify Local Structures in
  Atomic-Resolution Transmission Electron Microscopy Images}.
\newblock {\em Adv. Theory Simulations}, 1(8):1800037, aug 2018.
\newblock URL: \url{http://doi.wiley.com/10.1002/adts.201800037}, \href
  {http://arxiv.org/abs/1802.03008} {\path{arXiv:1802.03008}}, \href
  {https://doi.org/10.1002/adts.201800037} {\path{doi:10.1002/adts.201800037}}.

\bibitem{Ziatdinov2017}
Maxim Ziatdinov, Ondrej Dyck, Artem Maksov, Xufan Li, Xiahan Sang, Kai Xiao,
  Raymond~R. Unocic, Rama Vasudevan, Stephen Jesse, and Sergei~V. Kalinin.
\newblock {Deep Learning of Atomically Resolved Scanning Transmission Electron
  Microscopy Images: Chemical Identification and Tracking Local
  Transformations}.
\newblock {\em ACS Nano}, 11(12):12742--12752, dec 2017.
\newblock URL: \url{http://pubs.acs.org/doi/10.1021/acsnano.7b07504}, \href
  {https://doi.org/10.1021/acsnano.7b07504}
  {\path{doi:10.1021/acsnano.7b07504}}.

\bibitem{Trentino2021}
Alberto Trentino, Jacob Madsen, Andreas Mittelberger, Clemens Mangler, Toma
  Susi, Kimmo Mustonen, and Jani Kotakoski.
\newblock {Atomic-Level Structural Engineering of Graphene on a Mesoscopic
  Scale}.
\newblock {\em Nano Lett.}, 21(12):5179--5185, 2021.
\newblock \href {https://doi.org/10.1021/acs.nanolett.1c01214}
  {\path{doi:10.1021/acs.nanolett.1c01214}}.

\bibitem{Aguiar2020}
Jeffery~A. Aguiar, Matthew~L Gong, and Tolga Tasdizen.
\newblock {Crystallographic prediction from diffraction and chemistry data for
  higher throughput classification using machine learning}.
\newblock {\em Comput. Mater. Sci.}, 173(August):109409, feb 2020.
\newblock URL: \url{https://doi.org/10.1016/j.commatsci.2019.109409
  https://linkinghub.elsevier.com/retrieve/pii/S0927025619307086}, \href
  {https://doi.org/10.1016/j.commatsci.2019.109409}
  {\path{doi:10.1016/j.commatsci.2019.109409}}.

\bibitem{Rutter2019}
Erica~M. Rutter, John~H. Lagergren, and Kevin~B. Flores.
\newblock {A Convolutional Neural Network Method for Boundary Optimization
  Enables Few-Shot Learning for Biomedical Image Segmentation}.
\newblock In {\em Domain Adapt. Represent. Transf. Med. Image Learn. with Less
  Labels Imperfect Data}, pages 190--198. Springer, New York, 2019.
\newblock URL: \url{http://link.springer.com/10.1007/978-3-030-33391-1{\_}22},
  \href {https://doi.org/10.1007/978-3-030-33391-1_22}
  {\path{doi:10.1007/978-3-030-33391-1_22}}.

\bibitem{Finn2017}
Chelsea Finn, Pieter Abbeel, and Sergey Levine.
\newblock {Model-agnostic meta-learning for fast adaptation of deep networks}.
\newblock {\em 34th Int. Conf. Mach. Learn. ICML 2017}, 3:1856--1868, 2017.
\newblock \href {http://arxiv.org/abs/1703.03400} {\path{arXiv:1703.03400}}.

\bibitem{Altae-Tran2017}
Han Altae-Tran, Bharath Ramsundar, Aneesh~S. Pappu, and Vijay Pande.
\newblock {Low Data Drug Discovery with One-Shot Learning}.
\newblock {\em ACS Cent. Sci.}, 3(4):283--293, apr 2017.
\newblock URL: \url{https://pubs.acs.org/doi/10.1021/acscentsci.6b00367}, \href
  {http://arxiv.org/abs/1611.03199} {\path{arXiv:1611.03199}}, \href
  {https://doi.org/10.1021/acscentsci.6b00367}
  {\path{doi:10.1021/acscentsci.6b00367}}.

\bibitem{Yao2021}
Fupin Yao.
\newblock {Machine learning with limited data}.
\newblock 2021.
\newblock URL: \url{http://arxiv.org/abs/2101.11461}, \href
  {http://arxiv.org/abs/2101.11461} {\path{arXiv:2101.11461}}.

\bibitem{He2016b}
Kaiming He, Xiangyu Zhang, Shaoqing Ren, and Jian Sun.
\newblock {Deep Residual Learning for Image Recognition}.
\newblock In {\em 2016 IEEE Conf. Comput. Vis. Pattern Recognit.}, pages
  770--778. IEEE, jun 2016.
\newblock URL: \url{http://ieeexplore.ieee.org/document/7780459/}, \href
  {https://doi.org/10.1109/CVPR.2016.90} {\path{doi:10.1109/CVPR.2016.90}}.

\bibitem{Kaufmann2021}
Kevin Kaufmann, Hobson Lane, Xiao Liu, and Kenneth~S. Vecchio.
\newblock {Efficient few-shot machine learning for classification of EBSD
  patterns}.
\newblock {\em Sci. Rep.}, 11(1):8172, dec 2021.
\newblock URL: \url{https://doi.org/10.1038/s41598-021-87557-5
  http://www.nature.com/articles/s41598-021-87557-5}, \href
  {https://doi.org/10.1038/s41598-021-87557-5}
  {\path{doi:10.1038/s41598-021-87557-5}}.

\bibitem{Akers2021}
S.~Akers, E.~Kautz, A.~Trevino-Gavito, M.~Olszta, B.~Matthews, L.~Wang, Y.~Du,
  and S.R. Spurgeon.
\newblock {Rapid and Flexible Semantic Segmentation of Electron Microscopy Data
  Using Few-Shot Machine Learning}.
\newblock \href {https://doi.org/10.21203/rs.3.rs-346102}
  {\path{doi:10.21203/rs.3.rs-346102}}.

\bibitem{Doty2021}
Christina Doty, Shaun Gallagher, Wenqi Cui, Wenya Chen, Shweta Bhushan,
  Marjolein Oostrom, Sarah Akers, and Steven~R. Spurgeon.
\newblock {Design of a Graphical User Interface for Few-Shot Machine Learning
  Classification of Electron Microscopy Data}.
\newblock jul 2021.
\newblock URL: \url{http://arxiv.org/abs/2107.10387}, \href
  {http://arxiv.org/abs/2107.10387} {\path{arXiv:2107.10387}}.

\bibitem{Authors2021}
The~ZeroMQ Authors.
\newblock {ZeroMQ: An open-source universal messaging library}, 2021.
\newblock URL: \url{https://zeromq.org}.

\bibitem{Laboratory2021}
Pacific Northwest~National Laboratory.
\newblock {About DataHub}, 2021.
\newblock URL: \url{https://data.pnnl.gov/about}.

\bibitem{Mitchell2005a}
D.R.G. Mitchell and B.~Schaffer.
\newblock {Scripting-customised microscopy tools for Digital Micrograph™}.
\newblock {\em Ultramicroscopy}, 103(4):319--332, jul 2005.
\newblock URL:
  \url{https://linkinghub.elsevier.com/retrieve/pii/S0304399105000355}, \href
  {https://doi.org/10.1016/j.ultramic.2005.02.003}
  {\path{doi:10.1016/j.ultramic.2005.02.003}}.

\bibitem{Gatan2021a}
Gatan.
\newblock {GMS Python}, 2021.
\newblock URL: \url{https://www.gatan.com/python-installation}.

\bibitem{Gong2020a}
Yongshuai Gong, Yiman Dong, Biao Zhao, Runnan Yu, Siqian Hu, and Zhan'Ao Tan.
\newblock {Diverse applications of MoO 3 for high performance organic
  photovoltaics: fundamentals, processes and optimization strategies}.
\newblock {\em J. Mater. Chem. A}, 8(3):978--1009, 2020.
\newblock URL: \url{http://xlink.rsc.org/?DOI=C9TA12005J}, \href
  {https://doi.org/10.1039/C9TA12005J} {\path{doi:10.1039/C9TA12005J}}.

\bibitem{Zollfrank2012}
Cordt Zollfrank, Kai Gutbrod, Peter Wechsler, and Josef~Peter Guggenbichler.
\newblock {Antimicrobial activity of transition metal acid MoO3 prevents
  microbial growth on material surfaces}.
\newblock {\em Mater. Sci. Eng. C}, 32(1):47--54, jan 2012.
\newblock URL: \url{http://dx.doi.org/10.1016/j.msec.2011.09.010
  https://linkinghub.elsevier.com/retrieve/pii/S092849311100258X}, \href
  {https://doi.org/10.1016/j.msec.2011.09.010}
  {\path{doi:10.1016/j.msec.2011.09.010}}.

\bibitem{Lyon2010}
S.B. Lyon.
\newblock {Corrosion of Molybdenum and its Alloys}.
\newblock In {\em Shreir's Corros.}, pages 2157--2167. Elsevier, 2010.
\newblock URL:
  \url{https://linkinghub.elsevier.com/retrieve/pii/B9780444527875001062},
  \href {https://doi.org/10.1016/B978-044452787-5.00106-2}
  {\path{doi:10.1016/B978-044452787-5.00106-2}}.

\bibitem{Nakahara1992}
S.~Nakahara and A.G. Cullis.
\newblock {Simple method for determining the absolute sense of image rotation
  in a transmission electron microscope}.
\newblock {\em Ultramicroscopy}, 45(3-4):365--370, nov 1992.
\newblock URL:
  \url{https://linkinghub.elsevier.com/retrieve/pii/030439919290148D}, \href
  {https://doi.org/10.1016/0304-3991(92)90148-D}
  {\path{doi:10.1016/0304-3991(92)90148-D}}.

\bibitem{schneider_2012}
Caroline~A. Schneider, Wayne~S. Rasband, and Kevin~W. Elicieri.
\newblock Nih image to imagej: 25 years of image analysis.
\newblock {\em Nature Methods}, 9(7):671--675, 2012.
\newblock \href {https://doi.org/https://doi.org/10.1038/nmeth.2089}
  {\path{doi:https://doi.org/10.1038/nmeth.2089}}.

\bibitem{rueden_2017}
Curtis~T. Rueden, Johannes Schindelin, Mark~C. Hiner, Barry~E. DeZonia,
  Alison~E. Walter, Ellen~T. Arena, and Kevin~W. Eliceiri.
\newblock Imagej2: Imagej for the next generation of scientific image data.
\newblock {\em BMC Bioinformatics}, 18(1), 2017.
\newblock \href {https://doi.org/https://doi.org/10.1186/s12859-017-1934-z}
  {\path{doi:https://doi.org/10.1186/s12859-017-1934-z}}.

\bibitem{schindelin_2012}
Johannes Schindelin, Ignacio Arganda-Carreras, Erwin Frise, Verena Kaynig, Mark
  Longair, Tobias Pietzsch, Stephan Preibisch, Curtis Rueden, Stephan Saalfeld,
  Benjamin Schmid, Jean-Yves Tinevez, Daniel~James White, Volker Hartenstein,
  Kevin Elicieri, Pavel Tomancak, and Albert Cardona.
\newblock Fiji: an open-source platform for biological-image analysis.
\newblock {\em Nature Methods}, 9(7):676--682, 2012.
\newblock \href {https://doi.org/https://doi.org/10.1038/nmeth.2019}
  {\path{doi:https://doi.org/10.1038/nmeth.2019}}.

\bibitem{2020SciPy-NMeth}
Pauli Virtanen, Ralf Gommers, Travis~E. Oliphant, Matt Haberland, Tyler Reddy,
  David Cournapeau, Evgeni Burovski, Pearu Peterson, Warren Weckesser, Jonathan
  Bright, St{\'e}fan~J. {van der Walt}, Matthew Brett, Joshua Wilson, K.~Jarrod
  Millman, Nikolay Mayorov, Andrew R.~J. Nelson, Eric Jones, Robert Kern, Eric
  Larson, C~J Carey, {\.I}lhan Polat, Yu~Feng, Eric~W. Moore, Jake
  {VanderPlas}, Denis Laxalde, Josef Perktold, Robert Cimrman, Ian Henriksen,
  E.~A. Quintero, Charles~R. Harris, Anne~M. Archibald, Ant{\^o}nio~H. Ribeiro,
  Fabian Pedregosa, Paul {van Mulbregt}, and {SciPy 1.0 Contributors}.
\newblock {{SciPy} 1.0: Fundamental Algorithms for Scientific Computing in
  Python}.
\newblock {\em Nature Methods}, 17:261--272, 2020.
\newblock \href {https://doi.org/10.1038/s41592-019-0686-2}
  {\path{doi:10.1038/s41592-019-0686-2}}.

\end{thebibliography}
\bibliographystyle{unsrturl}

\end{document}